\documentclass[journal]{IEEEtran}
\usepackage{subfigure}
\usepackage{setspace}
\usepackage{graphicx}
\usepackage{amsmath}
\usepackage{amssymb}
\usepackage{amsfonts}
\usepackage{epstopdf}
\usepackage{booktabs}
\usepackage{url}
\usepackage{algorithm}
\usepackage{algorithmicx}
\usepackage{multicol}
\usepackage{booktabs}
\usepackage{caption}	
\usepackage{color}
\usepackage{enumerate}

\newcommand*{\RR}[1]{\textcolor{magenta}{#1}}
\newcommand*{\my}[1]{\textcolor{black}{#1}}	
\begin{document}
%
\title{SimpleTrack:Adaptive Trajectory Compression with Deterministic Projection Matrix for Mobile Sensor Networks}
%
%
%

\author{Rajib~Rana,~\IEEEmembership{Member,~IEEE,}
        Mingrui~Yang,~\IEEEmembership{Member,~IEEE,}
        ~Tim~Wark,~\IEEEmembership{Senior Member,~IEEE,}
        ~Chun Tung~Chou,~\IEEEmembership{Member,~IEEE,}
       ~and~Wen~Hu,~\IEEEmembership{Senior Member,~IEEE}
\thanks{R. Rana is with the Department
of Computation Informatics, CSIRO (The Commonwealth Scientific and Industrial Research Organisation), Australia. e-mail: rajib.rana@csiro.au.}
\thanks{M. Yang, T. Wark and W. Hu are with the Department
of Computation Informatics, CSIRO. Chun Tung Chou is with the School of Computer Science and Engineering, UNSW, Sydney, Australia.}
}

\markboth{IEEE Sensors Journal}%
{Rana \MakeLowercase{\textit{et al.}}: A Deterministic Construction of Projection matrix for Adaptive Trajectory Compression}

\maketitle

\begin{abstract}
\RR{
Some mobile sensor network applications require the sensor nodes to transfer their trajectories to a data sink. This paper proposes an adaptive trajectory (lossy) compression algorithm based on compressive sensing. The algorithm has two innovative elements. First, we propose a method to compute a deterministic projection matrix from a learnt dictionary. Second, we propose a method for the mobile nodes to adaptively predict the number of projections needed based on the speed of the mobile nodes. Extensive evaluation of the proposed algorithm using 6 datasets shows that our proposed algorithm can achieve sub-metre accuracy. In addition, our method of computing projection matrices outperforms two existing methods. Finally, comparison of our algorithm against a state-of-the-art trajectory compression algorithm show that our algorithm can reduce the error by 10-60 cm for the same compression ratio. 
}
\end{abstract}


\IEEEpeerreviewmaketitle

\section{Introduction}
\label{sec:intro}
Mobile sensor networks (MSNs), which consists of autonomous embedded sensor nodes roaming freely, offer many new opportunities that are not available to their stationary counterparts. The Virtual Fencing (VF) \cite{bishop2007virtual} project that is being conducted out in our laboratory is one such example. A cattle farm generally covers an enormous area and it is costly to build fences around it. VF offers an alternative where no physical fencing is needed. The cattle carry an embedded device with GPS on board. The device constantly monitors the cow's location and if a cow tries to leave the farm, the device sends a stimulus (either an auditory or mild electric shock) to signal the cow to turn back. Ethical considerations are critical to the VF application. Locations of the animals and records of stimuli applied must be kept. This requires the device to record the \emph{trajectory} of the animal. Given the limited storage capacity of the embedded device, these stored trajectories must be uploaded to some server at some time. Due to the enormous size of the farm, base stations can only be installed at certain places. Therefore VF operates as a delay tolerant network. When a cow is getting close to a base station, the device on the animal makes use of the short transmission opportunity available to transfer the stored trajectory to the server. Given this limited transmission opportunity, as well as limited storage on the device, \emph{trajectory compression} is important. 

Data compression is a richly researched field, with many well known algorithms such as Lempel-Ziv \cite{cover2012elements} and many others. However, in the context of trajectory compression in MSNs, we also need to take the limited computation and transmission resources of MSN nodes into consideration. This demands for us to look for simple compression algorithm with good space savings. The recently developed theory of Compressive Sensing (CS)~\cite{candes_robust,Donoho:06} offers such a possibility because its compression step is very simple. In fact, we demonstrated in our earlier work~\cite{Rana:2011:AAC:1966251.1966255} that such type of compression is feasible on an 8-bit Atmel Amega 1281 microcontroller with 8 kB RAM. However, we could only achieve an accuracy in the order of metres for the reconstructed trajectories in~\cite{Rana:2011:AAC:1966251.1966255}. In this paper, we propose an improved compression scheme SimpleTrack which achieves a sub-metre accuracy. 

SimpleTrack is also based on CS. Given a $n$-dimensional data vector $x$ to be compressed, SimpleTrack uses a $m \times n$ projection matrix $\Phi$ to compute compressed data vector $y = \Phi x$. This projection matrix $\Phi$ is fat (which means $m < n$), therefore the output of the compression step $y$ has a lower dimension compared with the original data vector $x$. The number of projections $m$ determines the size of the compressed data $y$. A smaller $m$ means lower storage and transmission requirement. The compressed vector $y$ is used during decompression to reconstruct an approximation of $x$. This reconstruction step requires the projection matrix $\Phi$ as well as a reconstruction basis $\Psi$. In order to achieve accurate reconstruction with low resource consumption, we need to make good choices of the three parameters: number of projections $m$, projection matrix $\Phi$ and reconstruction basis $\Psi$. In this paper, we make the following contributions:
\begin{itemize}

\item \RR{We propose an adaptive method, based on support vector regression (SVR) \cite{svr}, to enable the MSN nodes to dynamically choose the number of projections $m$ based on the their speed. 
}
\item \RR{We show that better reconstruction accuracy can be achieved by using a learnt dictionary together with a projection matrix computed from the dictionary. 
}

\item \RR{We propose a new method to compute projection matrix from a dictionary. Experimental results show that our proposed method outperforms two existing methods.  We also provide an explanation of why our proposed method works better.}
\item \RR{We perform extensive evaluation by using 6 datasets. 
The results show that our method is 10-60cm more accurate than the state-of-the-art trajectory compression algorithm SQUISH \cite{muckell2011squish}. 
}
\end{itemize}

This paper is organized as follows. Section ~\ref{sec:backGround} presents the background on CS and dictionary learning. We describe our proposed trajectory compression algorithm SimpleTrack in Section \ref{sec:st}. Evaluations of SimpleTrack are presented in Section~\ref{sec:eval}. Section~\ref{sec:related} discusses related work and Section~\ref{sec:conclusion} concludes the paper.   

%

\section{Background}
\label{sec:backGround}
SimpleTrack uses CS and dictionary learning. We present an overview of these two topics in this section. 

\subsection{Compressive Sensing (CS)}
\label{sec:CS}

CS has received a lot of attention in the past decade because it can significantly reduce the sampling requirement in many applications~(\cite{sivapalan2011compressive,chew2012sparse,xu2011dynamic}). CS has also been used in wireless sensor networks (WSNs) for reducing the energy consumption in data gathering (\cite{Luo:2009wy,chou2009energy,rana2010energy,shen2013nonuniform,wei2012distributed}) and the computational requirements on sensors (\cite{shen2012efficient,Chou:2013ch,shen2011non}). We will review the aspects of CS which are necessary for understanding this paper, more details can be found in (\cite{Donoho:06}). 

Given a vector $x \in \mathbb{R}^n$, we can compute its representation $\theta \in \mathbb{R}^n$ in a basis $\Psi \in \mathbb{R}^{n \times n}$ by solving the linear equation
\begin{equation}
x = \Psi \theta
\label{eqn:psi_theta} 
\end{equation}
The representation $\theta$ is said to be \emph{compressible} if $\theta$ has a large number of elements with small magnitude. We can realise compression by setting these elements with small magnitude to zero. This can reduce the storage requirements. If we can find a basis in which a given vector $x$ has a compressible representation, we will also say that $x$ is compressible. The theory of CS applies to compressible vectors. 

CS considers the problem of recovering an unknown compressible vector $x$ from its projections. Let $\Phi$ be a ${m \times n}$ projection matrix with $m < n$. Consider the equation:
\begin{equation}
y = \Phi x + z 
\label{eqn:proj}
\end{equation}
where $z \in \mathbb{R}^n$ is a noise vector whose norm is bounded by $\epsilon$. CS aims to recover (or reconstruct) $x$ from $y$ and $\Phi$ given the knowledge that $x$ is compressible in the basis $\Psi$. CS shows that under certain conditions it is possible to recover $x$ by solving the following $\ell_1$ optimisation problem:  
\begin{equation}
     \min_{\hat{\theta}\in \mathbb{R}^n}\|\hat{\theta}\|_1 \quad \text{subject to } \|y - \Phi \Psi \hat{\theta}\|_2 \leq \epsilon.
     \label{eqn:l1_min}
 \end{equation}
Given $\hat{\theta}$, we can get an estimate of $x$ from $\hat{x} = \Psi \hat{\theta}$. 

In the context of trajectory compression, $x$ is the trajectory measured by an MSN node. The dimension of $x$ is large. The MSN node computes $y = \Phi x$ and transmits $y$ to the server. The server can compute an estimated trajectory $\hat{x}$ by using $y$, $\Phi$ and $\Psi$, by solving \eqref{eqn:l1_min}. Note that the compression is lossy with $1-\frac{m}{n}$ represents both space savings and reduction in wireless transmission requirement. 

The reconstruction error $\| \hat{x} - x \|$ depends on a number of factors. The number of projections $m$ must be large enough. A larger $m$ generally reduces the reconstruction error but increases computation and transmission requirements at the sensor. The theory of CS shows that the number of projections $m$ needed depends on the \emph{compressibility} of the vector $x$ in the basis $\Psi$. We say that a vector $x$ is more compressible if its representation has fewer number of dominant elements, i.e. fewer non-zero elements with large magnitude. The theory of CS shows that a smaller $m$ is needed if $x$ is more compressible. We will propose an adaptive method to determine $m$ on the MSN nodes in Section \ref{sec:st}. 

The choice of basis $\Psi$ also affects the reconstruction error. A basic requirement is that $x$ has to be compressible in the basis $\Psi$. In our previous work~\cite{Rana:2011:AAC:1966251.1966255}, we used standard bases for $\Psi$. In this paper, we show that learnt dictionary gives better performance. 

Another parameter that determines the achievable reconstruction error is the projection matrix $\Phi$. The requirements on $\Phi$ to achieve low reconstruction error is expressed in terms of sensing matrix $A = \Phi \Psi$. Two requirements have been stated in the literature, in terms of Restricted Isometry Property (RIP) \cite{candes2008restricted} and coherence~\cite{Candes2008:Introduction}. We will discuss coherence. The mutual coherence $\mu(A)$ for the sensing matrix $A$ is defined as
\begin{equation}
	\mu(A) = \max_{i < j} \frac{|\langle a_i, a_j \rangle|}{\|a_i\|_2 \|a_j\|_2},
\end{equation}
where $a_i$ and $a_j$ are columns of $A$. In words, the coherence measures the largest correlation between any two columns of $A$.  If $A$ contains \my{highly} correlated columns, the coherence is large. Otherwise, it is small. The role of the coherence is straightforward: the smaller the coherence is, the fewer number of projections $m$ is needed for low reconstruction error. We will present a new method of computing $\Phi$ in Section \ref{sec:st}. 

\subsection{Learning Sparsifying Dictionary}
\label{sec:sparseDomain}
\label{sec:SparseCodingForTrajectoryCompression}

In order that we can achieve good trajectory compression using CS, we need compressible representation of the trajectory in a certain basis. A method to obtain good basis is through \emph{learning sparisfying dictionary}. The terms \emph{sparse} and compressible are closely related. A vector $x$ is said to be sparse in the basis $\Psi$ if its representation in $\Psi$ has few non-zero elements. One can therefore view sparse vectors as a special case of compressible vectors. 

In dictionary learning, we assume that we are given a set of $P$ vectors $X = \{x_1, x_2, ..., x_P \}$ from $\mathbb{R}^n$ and we want to find a dictionary $D \in \mathbb{R}^{n \times d}$ such that the vectors $x_i$ are \emph{simultaneously} sparse in this dictionary $D$. In other words, if we compute the representations $\theta_i$ of $x_i$ via $x_i =D \theta_i$, we want all the vectors $\theta_i$ from $i = 1, ..., P$ to be sparse in $D$. 

In the context of trajectory compression, $X$ is a set of trajectories that we can use to learn a dictionary $D$. If a new trajectory $x$ is similar to those in $X$, then we expect that $x$ will have a fairly compressible representation in $D$. Since the more compressible $x$ is, the lower the requirement on the number of projections $m$ is, we can achieve the same reconstruction error with a smaller $m$, thus reducing wireless transmission requirement. 

Dictionary learning can be formulated as an optimisation problem. Given the set $ \{x_1, x_2, ..., x_P \}$, we want to find $n \times d$ dictionary matrix $D$ and coefficients $s_i \in \mathbb{R}^d$ such that
$x_i = D s_i \mbox{ for } i = 1, ..., P$
and $s_i$ are sparse. The requirement that $s_i$ is sparse can be imposed by forcing the vectors $s_i$ to have small $\ell_0$ norm since the $\ell_0$ norm of a vector counts the number of non-zero elements in the vector. However, using $\ell_0$ norm makes the optimisation problem hard, an alternative it to impose that the $\ell_1$ norm of $s_i$ be small. Let $d_j$ denote the $j$-th column of $D$. We define the optimisation problem 
\begin{eqnarray}
\min_{s_i,D} \sum_{i = 1}^P (\frac{1}{2}||x_i - D s_i||^2_2 + \lambda ||s_i||_1)  \nonumber\\
s.t. ||d_j||^2 \leq 1 \forall_j = 1,..,n	
\label{eqn:sparse_coding_new}
\end{eqnarray} 

The optimization problem~\eqref{eqn:sparse_coding_new} is convex with respect to each of the variables $D$ and $\{s_i\}$, when the other one is fixed. 
Therefore, practically it can be solved in two steps: learning the sparse coefficients $s_i$ keeping the dictionary $D$ fixed, and then learning the dictionary $D$ keeping the coefficients $s_i$ fixed~\cite{Mairal:2009:ODL:1553374.1553463,lee07}.  We use \my{SPAMS}~\cite{Mairal:2009:ODL:1553374.1553463} to solve \eqref{eqn:sparse_coding_new}. For dictionary learning, \my{SPAMS} \my{uses} the LARS-Lasso algorithm~\cite{efron2004least}, which is a homotopy method~\cite{Osborne99anew} providing the solutions for all possible values of $\lambda$. We choose SPAMS because it 
uses block-coordinate descent with warm restarts~\cite{Bertsekas99}, which 
guarantees the convergence to a global optimum. 

\section{SimpleTrack}
\label{sec:st}
This section describes our proposed trajectory compression algorithm SimpleTrack. The algorithm is based on CS and assumes that a number of trajectories are available for training. SimpleTrack has three parameters: number of projections $m$, dictionary $D$ and projection matrix $\Phi$. 
SimpleTrack uses dictionary learning to obtain $D$ from the training trajectories. We assume that $D$ has been computed in this section. 

In this section, we present a new method to compute projection matrix $\Phi$ from the learnt dictionary $D$ in Section \ref{sec:mTermSVD}. We will show that the projection matrix obtained from our method gives better performance than other methods in Section \ref{sec:eval}. In order to reduce the computation and transmission requirements of MSN nodes, we propose a method to adaptively determine the number of projections $m$ needed. This is based SVR and will be presented in Section \ref{sec:adaptiveCompression}. 

\subsection{Deterministic Construction of Projection Matrix}
\label{sec:mTermSVD}
We know from the theory of CS that a good choice of projection matrix $\Phi$ can reduce the reconstruction error. The projection matrix $\Phi$ should ideally be \emph{uncorrelated} with the dictionary $D$ (resp. the basis $\Psi$) such that the sensing matrix $A = \Phi D$ ($A = \Phi \Psi$) has low coherence. The problem of constructing good projection matrix has been considered in \cite{elad2007optimized} and \cite{Carin:12}. Our work is built on \cite{Carin:12} and we will show that our proposed method performs better than those in \cite{elad2007optimized} and \cite{Carin:12} in Section \ref{sec:eval}. 

We first describe the projection matrix construction method in \cite{Carin:12}. The method assumes that the dictionary $D \in \mathbb{R}^{n \times d}$ and the number of projections $m$ are the inputs. The method first computes the singular value decomposition (SVD) of $D$:
\begin{equation}
D = U \Lambda V^T 
\label{sec:svdD}
\end{equation}
where  
where $^T$ denotes matrix transpose, $\Lambda \in \mathbb{R}^{n\times d}$ contains the singular values in its main diagonal, and $U \in \mathbb{R}^{n \times n}$ and $V \in \mathbb{R}^{d\times d}$ are orthonormal matrices. The method in \cite{Carin:12} is to \emph{randomly} choose $m$ columns from the matrix $U$. Let $\tilde{U}_m$ be a $n \times m$ matrix formed by these $m$ randomly chosen columns from $U$. The method is to use $ \tilde{U}_m^T$ as the projection matrix. The rationale of the method is that the columns in $U$ are highly uncorrelated with the dictionary $D$, therefore the sensing matrix $\Psi D = \tilde{U}_m^T D$ will have low coherence. 

In our proposed method, we choose the $m$ columns in $U$ corresponding to the \emph{largest} $m$ singular values of $D$. We now explain why this is a better choice. To simplify notation, we assume that the SVD in \eqref{sec:svdD} has been permuted so that singular values appear in non-increasing order in the diagonal of $\Lambda$. With this notation, let $U_m$ denotes the sub-matrix containing the left-most $m$ columns of $U$; note that these $m$ columns correspond to the largest $m$ singular values of $D$. Our choice of projection matrix is therefore $U_m^T$. 

To understand why $U_m$ is a better choice, note that the trajectory reconstruction problem can be stated as estimating the unknown coefficient vector $s$ from the projection $y$ by solving 
$y = \Phi D s$.
We assume that the unknown coefficient vector $s$ comes from some probability distribution such that $\mathbb{E}[ s s^T] = \mathbb{I}$ where $\mathbb{E}$ and $\mathbb{I}$ denote respectively the expectation operator and the identity matrix. 
It can be shown that the mean signal power $\mathbb{E}[y^T y]$ can be written as:
\begin{align}
\mathbb{E}[y^T y] = trace(\Phi U \Lambda^2 U^T \Phi^T)
\end{align} 
If we impose the constraint that each row of the projection matrix $\Phi$ has unit norm, then the $\Phi$ that maximizes $\mathbb{E}[y^T y]$ is given by the first $m$ rows of $U^T$ (or $U_m^T$), i.e. the $m$ left singular vectors corresponding to the largest $m$ singular values. This shows that our choice of projection matrix maximises the signal power of $y$. A higher signal power typically translates to lower estimation error. 
We will show, through numerical evaluations in Section \ref{sec:eval}, that our method of computing projection matrix has two advantages.
\begin{enumerate}
\item For a given number of projections $m$, it reduces the trajectory reconstruction error.
\item Because we use a deterministic way of computing the projection matrix, we get less variability in the error of trajectory reconstruction.
\end{enumerate} 

\subsection{Adaptive Compression} 
\label{sec:adaptiveCompression}

The number of projections $m$ is an important parameter in SimpleTrack because it controls the trade-off between the computation/transmission requirements at the MSN nodes and the reconstruction accuracy of the trajectory. For a given reconstruction error, we want to use the smallest $m$ required. If the MSN nodes were powerful enough, they could compute the projection $y$ for different choices of $m$ and then perform the reconstruction to determine the exact reconstruction error; this would allow them to choose the best $m$ for a given reconstruction error. However, this is not feasible for MSN nodes. In fact, we require a simple way for the MSN nodes to choose $m$. 

Our proposal is to use off-line learning based on the hypothesis that number of projections required is correlated with the speed of movement. This is fairly intuitive. Low speed means little movement or little information in the trajectory, and vice versa. Therefore, we investigate the use of some function of speed to predict the number of projections required. Our off-line learning method requires a set of training trajectories $\{ x_1, x_2, ..., x_P \}$, a parameter $m_{\max}$ which is the maximum number of projections needed by the node and a parameter $\xi$ which is the maximum trajectory reconstruction error that we can tolerate. We first perform the following:
\begin{enumerate}
\item Use the training trajectories to compute the dictionary $D$ using the method in Section~\ref{sec:sparseDomain}.  
\item Use $D$ and $m_{\max}$ to compute a projection matrix $\Phi$ with $m_{\max}$ rows using the method in Section~\ref{sec:mTermSVD}.  
\item For each trajectory $x_i$, 
\begin{enumerate} 
\item Determine $s_i$ which is a statistics of the speeds within the trajectory. 
\item Determine the smallest number of projections $m_i$ required so that the reconstruction error is less than the tolerance $\xi$. 
\end{enumerate}
\end{enumerate} 

Note that for step 3a, there are many possible choices of statistics, e.g. mean or median speed. We will examine different choices in Section \ref{sec:eval}. After performing the above calculations, we have the training set $\{(s_1,m_1),...,$ $(s_P,m_P)\}$. We use $\epsilon$-SVR to determine a function $m = g(s)$ where $s$ is the speed statistics and $m$ is the number of projections that should be used for $s$. The parameter $\epsilon$ in $\epsilon$-SVR can be used to control the accuracy of the fit. The function $m = g(s)$ can be implemented on the MSN nodes using a look-up table, see ~\cite{Rana:2011:AAC:1966251.1966255}. We use the matlab library \texttt{LIBSVM}~\cite{CC01a} to implement $\epsilon$-SVR. 
We choose radial basis function (RBF) kernel, rather than linear kernel, because it is more suitable for small number of features. We found that the RBF kernel performs better than the linear kernel for our datasets.
\RR{
\subsection{Pseudocode for SimpleTrack Encoding and Decoding}
The pseudocode of the SimpleTrack encoding and decoding processes is shown in Algorithms~\ref{alg:simpleTrackEncoding} and~\ref{alg:simpleTrackDecoding} respectively. The encoding takes place on the MSN nodes while the decoding takes place at the sink. The encoding process turns an input trajectory segment $x$ with $n$ data points into a $m$-vector $y$ with $m < n$, which is transmitted to the basestation. The decoding process uses $y$ to recover an approximation of $x$.}

\RR{
SimpleTrack requires three inputs that are computed offline: a look-up table, matrix $\tilde{U}$ and dictionary $D$. The look-up table is used to adaptively determine the number of projections $m$ required for each trajectory segment using a speed statistics of that segment, see Section \ref{sec:adaptiveCompression}. We will evaluate the performance of different choices of speed statistics in Section \ref{sec:speed:stats}.
We see from Section \ref{sec:mTermSVD} that if $m$ projections are required, then the projection matrix is formed by the left-most $m$ columns of $U$ where $U$ contains the left singular vectors arranged in descending order of singular values. The columns of $U$ have to be stored on the sensor nodes to enable them to compute the projections but we do not need all the columns in $U$. Let $m_{\max}$ be the maximum number of projections a sensor node has to perform. We extract the left-most $m_{\max}$ columns of $U$ and store them in $\tilde{U}$, which will be stored on the sensor nodes. Lastly, the dictionary $D$ is obtained using the process described in Section \ref{sec:sparseDomain}. Note that the dictionary $D$ is only needed by the sink for decoding. The sensor nodes do not require the dictionary, but they need to store the $m_{\max} \times n$ projection matrix $\tilde{U}$, which has a smaller dimension compared with dictionary $D$. Finally, note that the input data size $n$ is fixed for a given type of subjects (e.g. animal type), while the size of the encoding output $m$ is variable and is determined by the look-up table.
\begin{algorithm}[t]
\caption{SimpleTrack encoding at MSN nodes} 
\begin{algorithmic}[1] 
\State {\bf Inputs}: Trajectory segment $x$ with $n$ data points; Look-up table mapping speed to number of projections; matrix $\tilde{U}$
\State Determine speed statistics $s$ of trajectory segment $x$ 
\State Search the look-up table in the program memory to find the number of projections $m$ needed for speed statistics $s$
\State Extract the first $m$ columns of $\tilde{U}$, which is $\tilde{U}_m$
\State Compute projections $y =  \tilde{U}_m^T x$.
\State Transmit the $m$-vector $y$ and $m$ to the basestation when the node is in range 
\end{algorithmic}
\label{alg:simpleTrackEncoding}
\end{algorithm}
\begin{algorithm}[t]
\caption{SimpleTrack decoding at network sink} 
\begin{algorithmic}[1] 
\State {\bf Inputs}: Vector $y$; Number of projections $m$; Dictionary $D$; $n\times n$ Matrix $\tilde{U}$ 
\State // Matrix $\tilde{U}$ identical to that on sensor nodes
\State Extract the first $m$ columns of $\tilde{U}$, which is $\tilde{U}_m$. 
\State Set $\Phi = \tilde{U}_m^T$ and $\Psi = D$. Solves optimisation problem \eqref{eqn:l1_min} to obtain $\hat{\theta}$. 
\State {\bf Output}: Estimated trajectory is $\Psi \hat{\theta}$.
\end{algorithmic}
\label{alg:simpleTrackDecoding}
\end{algorithm}
}
\RR{
\subsection{Complexity Analysis} 
\label{sec:complexityAnalysis}
The SimpleTrack decoding takes place at the server with plentiful of resources. Therefore, this complexity analysis focuses on encoding which takes place at the sensor nodes. Table look-up and projection computation are the two key operations in encoding. Table look-up can be efficiently done by binary search. Our earlier experience in~\cite{Rana:2011:AAC:1966251.1966255} shows that only a small look-up table is required. Therefore the dominant computation cost is to calculate the projections.  The projection calculations require multiplying a $m \times n$ matrix with a $n$-vector which has a complexity of the order ${\cal O}(mn)$. Experimental results on computation time will be presented in Section \ref{sec:sysperf}. 
}

\section{Performance evaluation}
\label{sec:eval}

\subsection{Datasets}
Our motivation for studying trajectory compression comes from the VF application. We will use a dataset obtained from VF for performance evaluation. In order to show that our proposed trajectory compression method is general, we supplement the evaluation by using 5 publicly available datasets on pedestrian mobility. 

\subsubsection{Animal Dataset}
The data were collected from a cattle monitoring trial~\cite{bishop2007virtual} 
with 36 cows for 49 hours. The data was collected using a wireless platform (mounted on cow collar) with a Nordic
NRF905 transceiver. The platform was connected to a Ublox4 GPS receiver using the Serial Peripheral Interface. The GPS data was collected at a 2 Hz sampling rate.

\subsubsection{Pedestrian Dataset}
We use publicly available pedestrian mobility traces from the CRAWDAD data repository~\cite{ncsu-mobilitymodels-2009-07-23}. The datasets and the number of traces in each dataset are: 
\begin{enumerate}[(i)]
\item NCSU with 35 traces  
\item KAIST with 46 traces 
\item New York City with 30 traces 
\item North Carolina state fair with 8 traces 
\item Disney World (Orlando) with 15 traces  
\end{enumerate}
For all the traces in these datasets, positions were recorded by Garmin GPS 60CSx handheld receivers every 30 seconds. It is known that these GPS receivers are accurate to within 3 meters for 95\% of the time in North America. 
\RR{
\subsubsection{Speed Variability in the Dataset}
Figure~\ref{fig:speedHistogram} shows the speed of a moving pedestrian and an animal over time. It can be seen that the subjects move at varying speed over time. This observation also applies to other datasets. 
\begin{figure}
\centering
\subfigure[Pedestrian]{
\includegraphics[width = 0.45\linewidth]{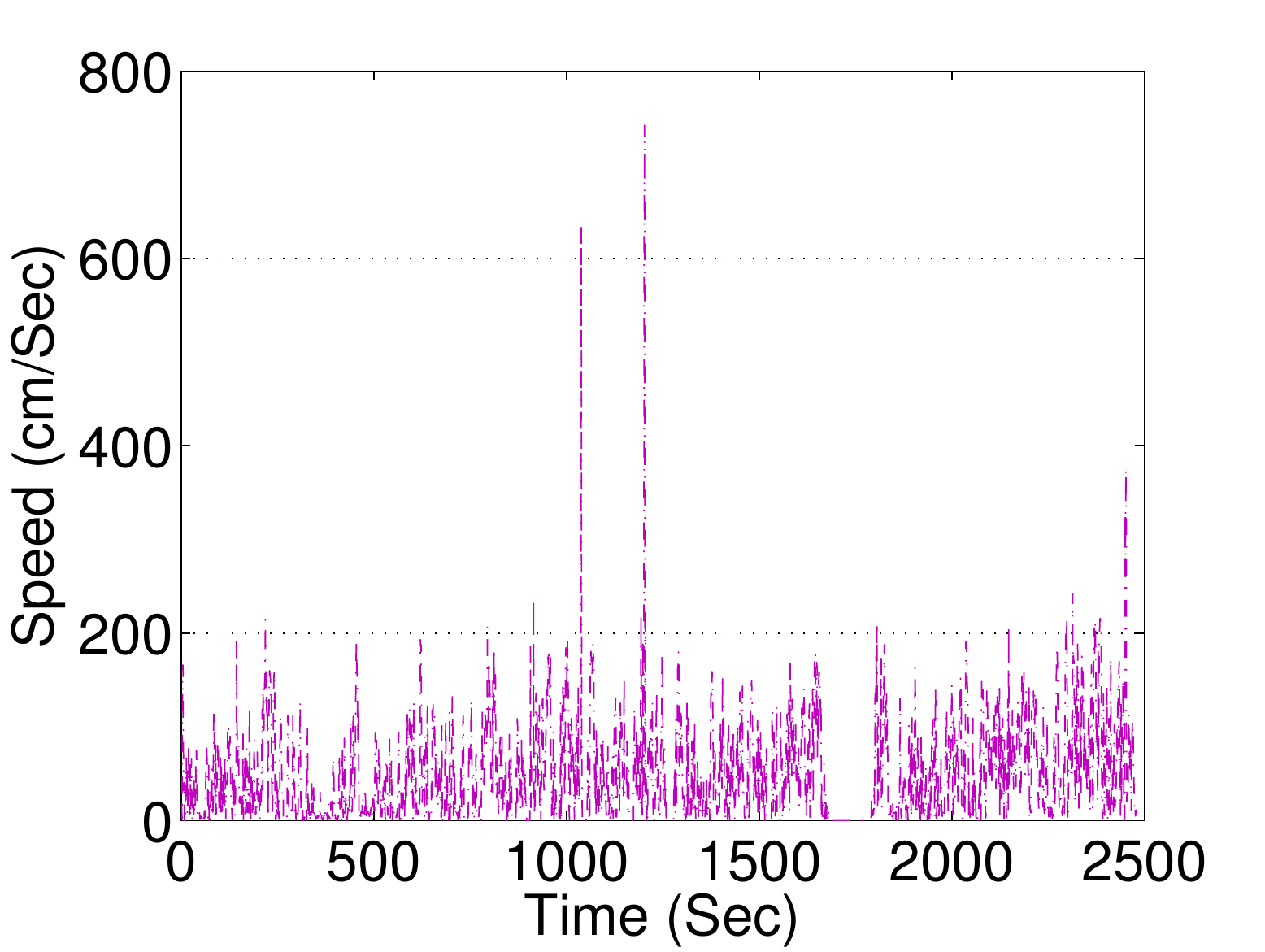}
}
\subfigure[Animal]{
\includegraphics[width=0.45\linewidth]{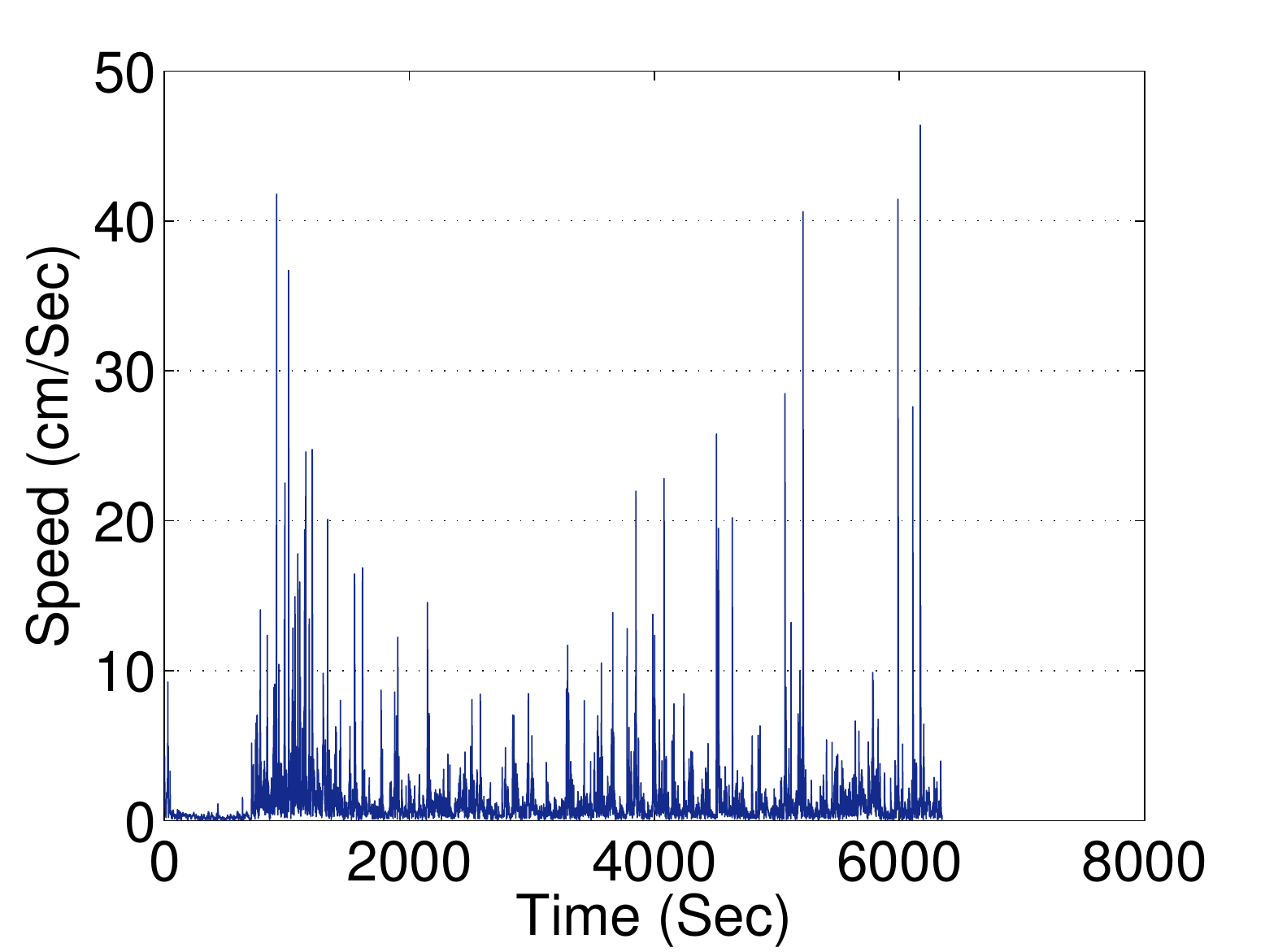}
}
\caption{\RR{Speed time series.}}
\label{fig:speedHistogram}
\end{figure}
}


\subsection{Methodology}
\subsubsection{Segment size and data format}
We divide each trace into segments consisting of a \RR{{\sl fixed}} number of data points. We determine the segment size using the speed. For cattle which do not move quickly, each segment consists of 128 data points or 64 seconds. For pedestrians, which have a higher average speed, each segment consists of 32 data points. Note that another reason for choosing a shorter segment for the pedestrian data is because of a longer sampling period of 30s. 

Each data point consists of two values: Easting (eastward-measured distance) and Northing (northward-measured distance). We encode/decode the easting and northing values in a segment separately. For the animal (resp. pedestrian) datasets, each trajectory segment gives two (resp. 32-) 128-vectors, one for easting and northing. 

\subsubsection{Preprocessing}
In order to consider only walking we filtered out segments where a speed of $6$km/hr was exceeded. All the measurements were converted into meters and the the data of the segment were subtracted by the mean of the segment.

\subsubsection{Training and test sets}
We used $80\%$ of the segments for training and the remaining $20\%$ for validation testing. For compression methods that require a dictionary, the training set is used to compute the dictionary. For the 5 pedestrian datasets, a separate dictionary is computed for each dataset.


\subsubsection{Performance metrics}
We measure the performance of each compression method by the accuracy of the reconstructed trajectories. Let $J$ be the number of segments in the test set. Let $N_{j,i}$ and $E_{j,i}$ denote, respectively, the Northing and Easting for the $i$-th data point in the $j$-th segment in the test set, with $1 \leq i \leq \frac{n}{2}$ and $1 \leq j \leq J$. We use $\hat{N}_{j,i}$ and $\hat{E}_{j,i}$ to denote the corresponding reconstructed Northing and Easting. We use the average distance between the original and reconstructed trajectories, Average Distance Error (ADE), to measure the reconstruction performance. ADE is defined as:
\begin{align}
\mbox{ADE} = \frac{2}{nJ}\sum_{j=1}^J\sum_{i=1}^{\frac{n}{2}} ({(N_{j,i}-\hat{N}_{j,i})^2+(E_{j,i}-\hat{E}_{j,i})^2})^{1/2}
\end{align}

\begin{figure*}
\centering
\includegraphics[width=0.9\linewidth]{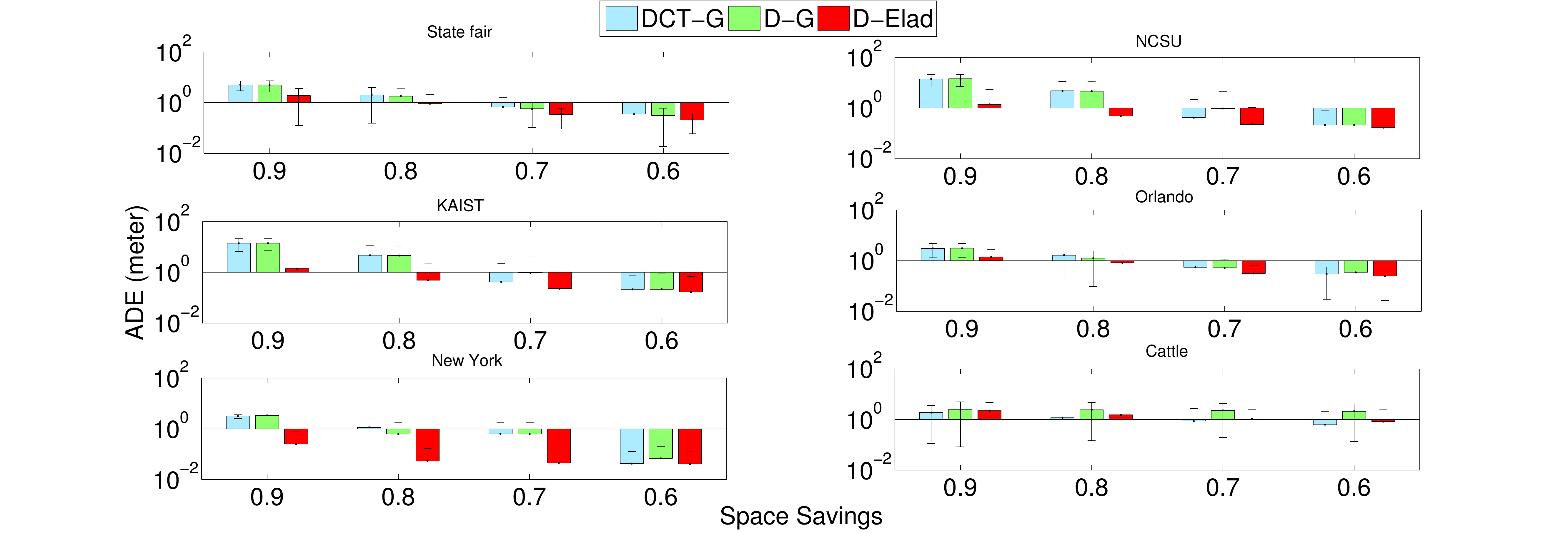}
\caption{\RR{ADE versus space savings. Methods: DCT-G, D-G and D-Elad}}
\label{fig:SpaceSavingsADE1}
\end{figure*}
\subsection{Compression methods}
In order to show that SimpleTrack gives good performance, we compare it against a number of methods. Two key ideas of SimpleTrack are to use a learnt dictionary and a new projection matrix computation method to improve trajectory compression. We therefore compare SimpleTrack against methods that do not use learnt dictionary and other methods of computing projection matrix. The compressive sensing based trajectory compression methods that we will study are: 
\begin{itemize}
\item \emph{Method DCT-G} We proposed a compressive sensing based trajectory compression method ~\cite{Rana:2011:AAC:1966251.1966255} which uses standard bases and random projection matrices. In ~\cite{Rana:2011:AAC:1966251.1966255}, we compared the compressibility of animal trajectories in different bases, including Discrete Cosine Transform (DCT) and various wavelet bases. We found that the animal trajectories are most compressible in the DCT basis. Multiple types of random projection matrices have been proposed in compressive sensing as the projection matrix, including Gaussian and Bernoulli random matrices. We found that difference projection matrices gave similar reconstruction performance. The trajectory compression method DCT-G is based on using DCT basis and Gaussian projection matrices. 
\item \emph{Method D-G} The D-G trajectory compression methods uses a learnt \emph{dictionary}, this is what the letter `D' stands for. The method uses random Gaussian matrices as projection matrices. 
\item \emph{Method D-SVDRandom} The D-SVDRandom trajectory compression method uses a learnt dictionary. This method uses the algorithm in \cite{Carin:12} to compute the projection matrix from the dictionary $D$. This method is based on choosing $m$ random left singular vectors from the SVD of $D$ and is discussed in section \ref{sec:mTermSVD}. The ``SVDRandom" part of the name is referring to this method of computing projection matrix.
\item \emph{Method D-Elad} Other methods for computing projection matrices from dictionaries can also be found in the literature. Elad~\cite{elad2007optimized} proposed a method to obtain an optimised projection matrix from a dictionary by minimising the mutual coherence. The D-Elad method uses a learnt dictionary and Elad's method of computing an optimised projection matrix. 
\item \emph{SimpleTrack} This is the method proposed in this paper. SimpleTrack uses a learnt dictionary and the projection matrix computation method proposed in Section \ref{sec:mTermSVD}. SimpleTrack also uses an adaptive method to determine the number of projections $m$ described in Section \ref{sec:adaptiveCompression}. 
\end{itemize}

Finally, we compare the compression performance of SimpleTrack against SQUISH~\cite{muckell2011squish}. SQUISH is a powerful GPS compression algorithm recently proposed by Muckell et al.~\cite{muckell2011squish}. It is a \RR{\sl non-dictionary} based method and works on the principles of synchronous Euclidian distance. Muckell et al. have shown that SQUISH performs better than other prominent trajectory compression methods such as Uniform Sampling, Online Dead Reckoning, and Online Douglas-Peucker. 

In this section, we study the reconstruction performance measured by ADE for the different compressive sensing based trajectory compression methods. We vary the space savings $1-\frac{m}{n}$ between 0.9 and 0.6 and calculate the ADE for each method. \RR{Note that space savings lower than 0.6 are not shown because their achievable ADE is comparable to that of 0.6 space savings.} For SimpleTrack, we disable the adaptive computation of $m$ and assume the value $m$ is given. 

\subsection{ADE for different space savings}

We first compare DCT-G, D-G and D-Elad to show that learnt dictionary and optimised projection matrix together give much better ADE for each space savings. Figure \ref{fig:SpaceSavingsADE1} shows the ADE versus space savings for the 6 datasets. The vertical axes of these plots show ADE measured in metres. The central horizontal lines are at the level of 1m for ADE. If a method achieves an ADE below 1m (i.e., sub-metre accuracy), the bar will appear below this line in the plots. Note that the vertical axes use logarithmic scale so that the magnitudes of sub-metre ADE can be distinguished. These results show that D-Elad outperforms DCT-G and D-G for all datasets and all space savings. This shows learnt dictionary, together with a good projection matrix, can improve the accuracy for a given $m$. We will now use D-Elad as the benchmark for the next study. 

\begin{figure*}
\centering
\includegraphics[width=0.9\linewidth]{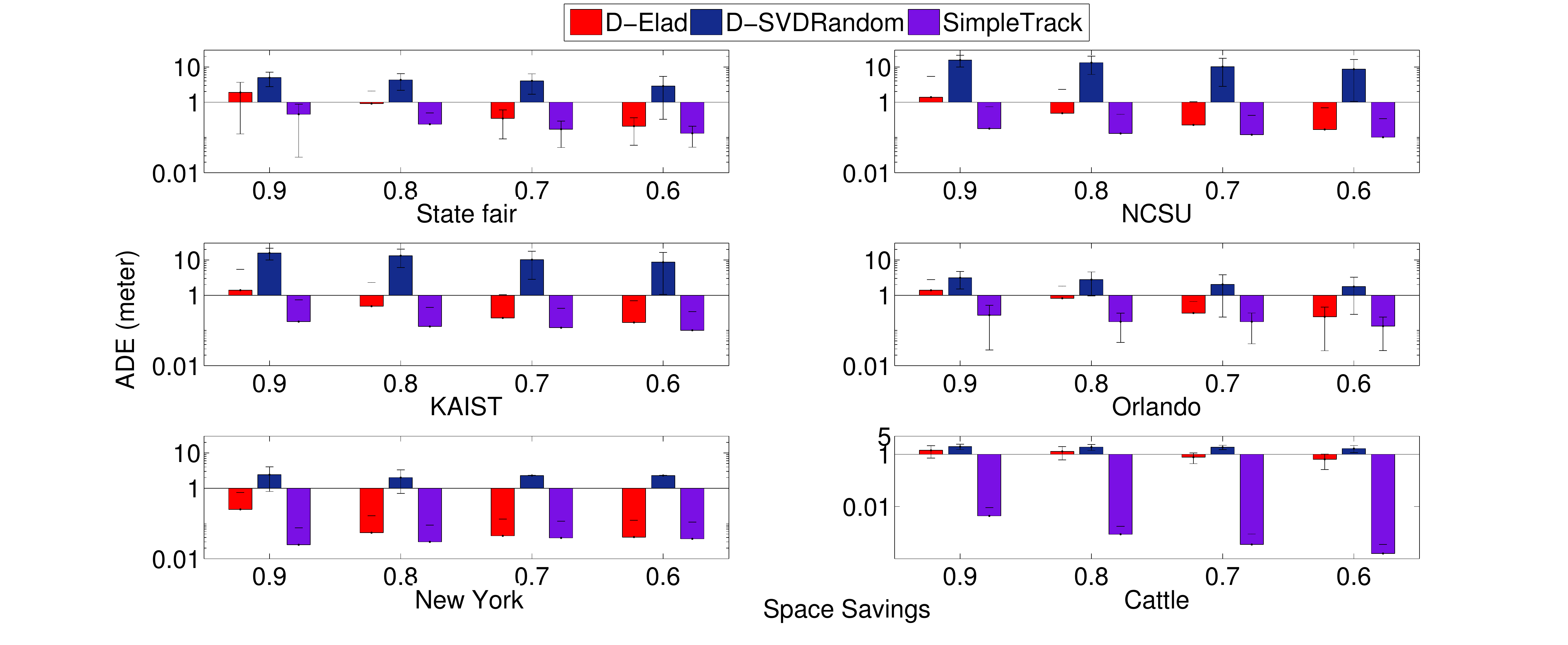}
\caption{\RR{ADE versus space savings. Methods: SimpleTrack, D-Elad and D-SVDRandom.}}
\label{fig:allTogetherSpaceSavingsADE}
\end{figure*}

In this study, we assume dictionary is used and we compare D-Elad, D-SVDRandom and SimpleTrack. Note that these three methods use the \emph{same} dictionary, the difference is the method to compute the projection matrix. The ADE versus space savings plots for the six datasets are shown in Figure \ref{fig:allTogetherSpaceSavingsADE}. It can be seen that SimpleTrack outperforms the other two methods for all datasets and space savings. In particular, SimpleTrack achieves sub-metre accuracy for all the space savings used. We can therefore conclude that: (1) SimpleTrack has the best performance compared to other compressive sensing based trajectory compression methods. (2) SimpleTrack can achieve sub-metre accuracy. (3) Our method of computing a projection matrix from a dictionary performs well. (4) The error bars also show our projection matrix reduces the variability in ADE. 

\begin{figure}
\centering
\includegraphics[width = 0.9\columnwidth]{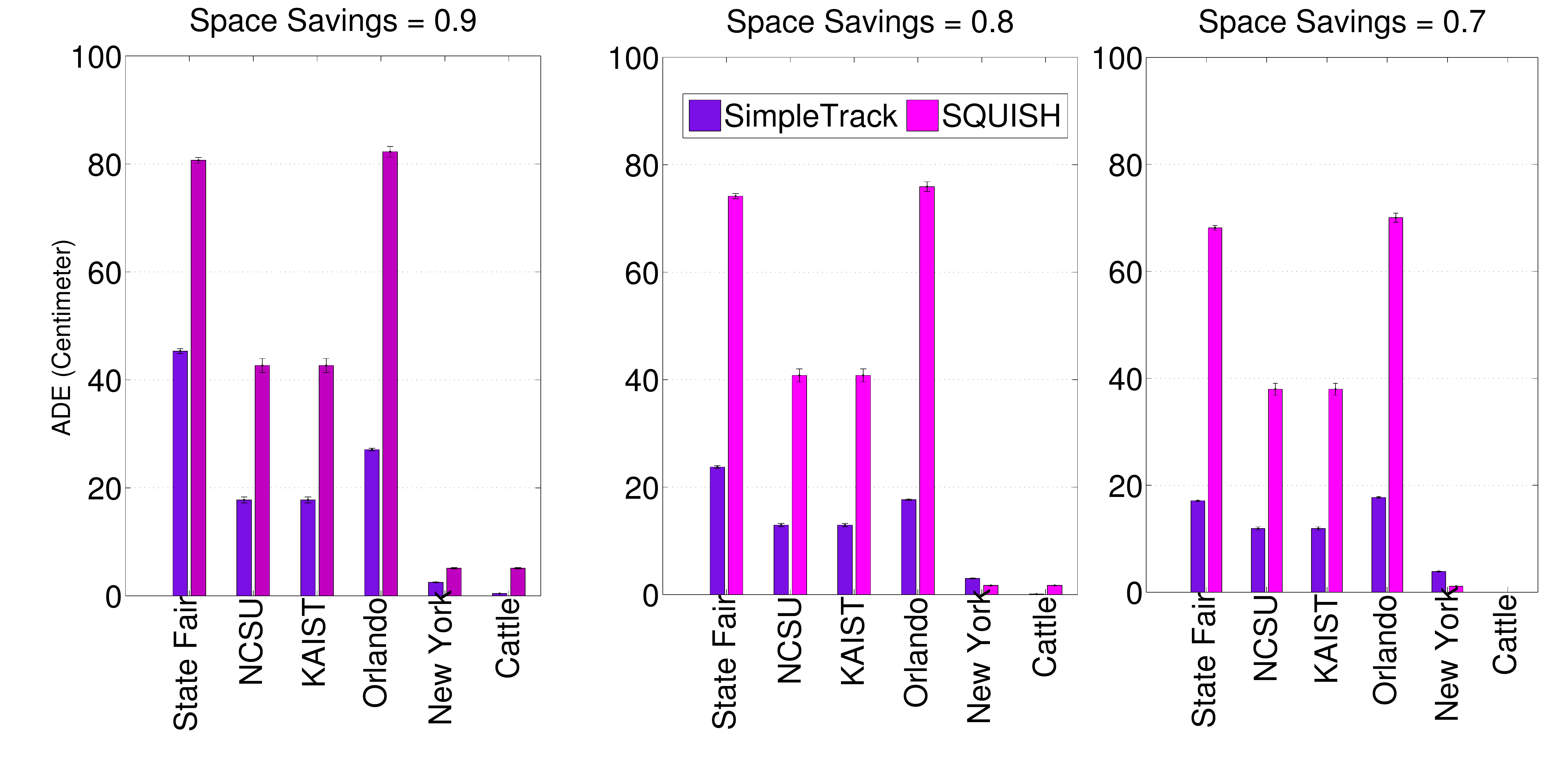}
\caption{SimpleTrack VS SQUISH.}
\label{fig:SQUISHvsUS}
\end{figure}

Finally, we contrast the performance of SimpleTrack with that of SQUISH in Figure \ref{fig:SQUISHvsUS}. For the cattle datasets\, SimpleTrack performs marginally better than SQUISH. For example, its reconstruction error is $10$ cm less than SQUISH. However, for the pedestrian datasets, SimpleTrack  performs much better. The reconstruction error can be $60$ cm less than SQUISH. 

\RR{The complexity of SimpleTrack is $O(mn)$, where $m$ is the number of projections and $n$ is the number of data points in the trajectory segment. SQUISH has a complexity $O(n\log{(\beta)})$, where $\beta$ is the size of the buffer. There is no indication on the buffer size is in the SQUISH paper~\cite{muckell2011squish}. We know from compressive sensing that $m \ll n$, therefore, it may be reasonable to assume that $m \sim \log{(\beta)}$. Thus, the complexity of SimpleTrack is comparable to that of SQUISH. In summary, SimpleTrack offers better accuracy while being similarly complex as SQUISH.}

\begin{table}[t]
\centering
\caption{Correlation between speed and number of measurements.}
\begin{tabular}{|c|c|c|}
\hline
&\multicolumn{2}{c|}{Correlation Coefficient}\\
\cline{2-3}
Speed &pedestrian& animal\\
parameter(s)&&\\\hline
mean&0.6&0.7\\\hline
variance&0.1& 0.6\\\hline
median&0.6& 0.6\\\hline
maximum&0.1&0.6\\\hline
minimum&0.2& 0.1\\\hline
\end{tabular}
\label{tab:lookup}
\end{table}

\subsection{Adaptive compression} 
\subsubsection{Determining suitable speed statistics} 
\label{sec:speed:stats}
We first determine which speed statistics to be used in prediction the number of projections $m$. We use 5 speed statistics: mean, variance, median, maximum and minimum. For each training segment, we compute these 5 speed statistics as well as the minimum number of projections $m$ needed to achieve a 50cm trajectory reconstruction error. In order to determine which speed statistics is the best, we compute the correlation between the speed statistics and the required $m$. The correlations are shown in Table~\ref{tab:lookup}. We will use mean speed as speed statistics since it has the highest correlation for both pedestrian and animal datasets. We then follow the SVR-based method in Section \ref{sec:adaptiveCompression} to determine $m$ as a function of mean speed. 



%



\begin{figure*}
\centering
\subfigure[ADE]{
\includegraphics[width = 0.3\linewidth]{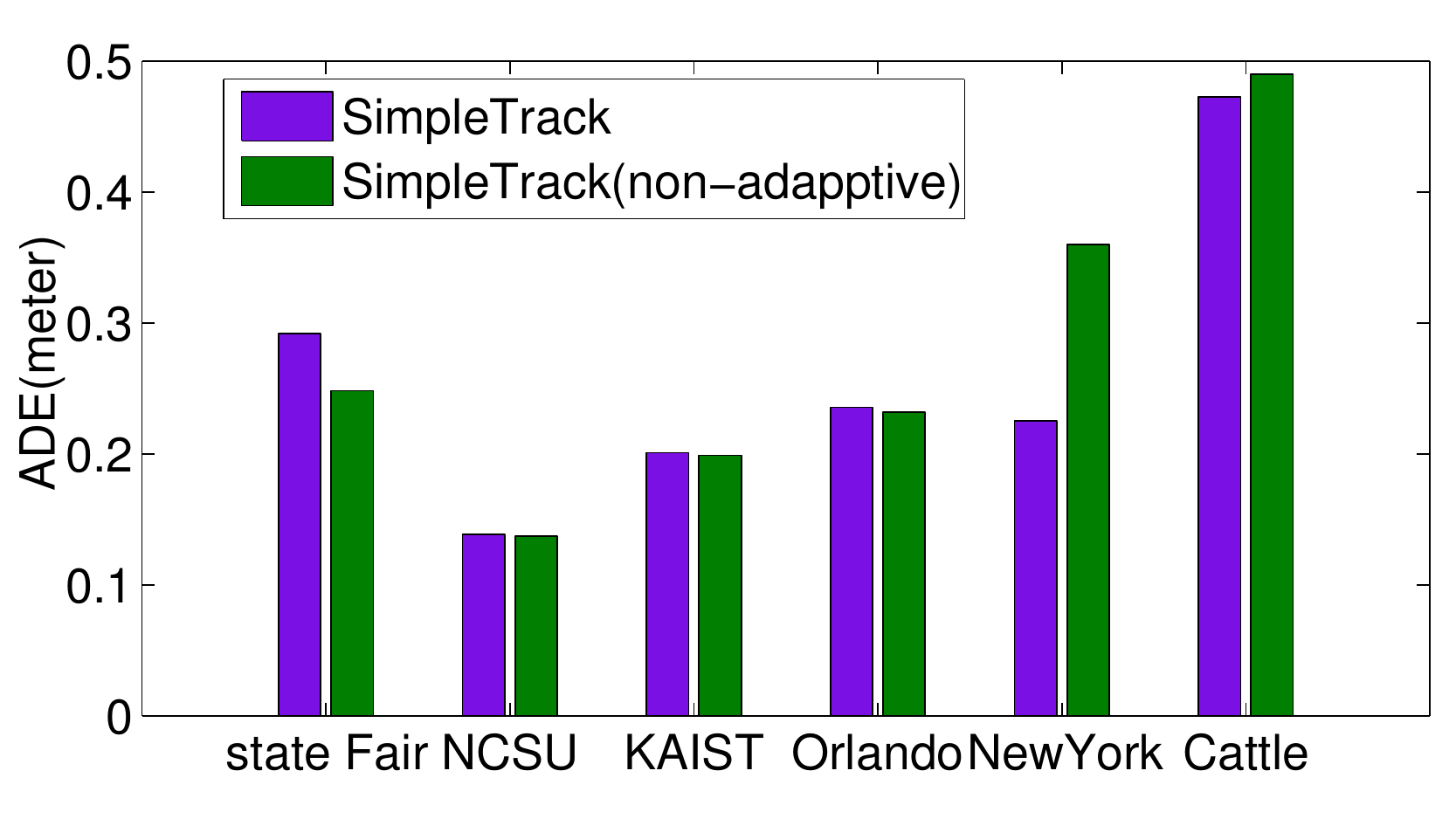}
\label{fig:errorComparison}
}
\subfigure[Prediction error (in \%) of number of projections needed to realise a 50cm reconstruction error.]{
\includegraphics[width = 0.3\linewidth]{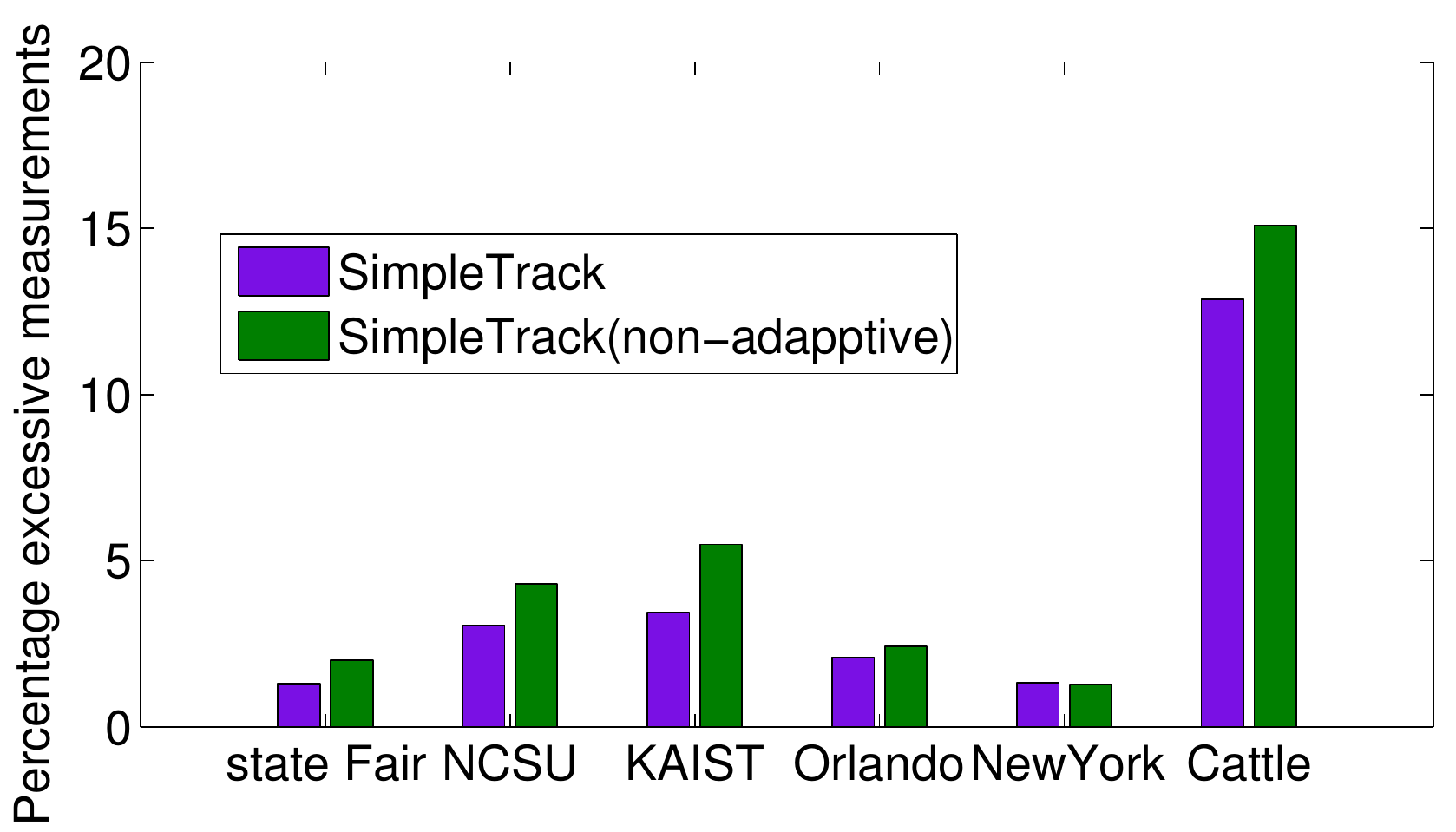}
\label{fig:percentageExcessiveMeasurements}
}
\caption{SimpleTrack versus SimpleTrack(non-adaptive).}
\label{fig:adaptiveVSnonAdaptive}
\end{figure*}



\subsubsection{Performance}
We compare SimpleTrack, which determines $m$ adaptively, against a non-adaptive version of SimpleTrack. We first use SimpleTrack on the test datasets and determine the mean number of projections $\bar{m}$ used by SimpleTrack for each test dataset. SimpleTrack (non-adaptive) uses a constant number of projections $\bar{m}$. Figure \ref{fig:errorComparison} compares the ADE realised by SimpleTrack and SimpleTrack (non-adaptive). It shows that SimpleTrack can achieve sub-metre accuracy and performs similar to SimpleTrack(non-adaptive), however, the latter requires the number of projections to be known beforehand, which is un-realistic. 

We also test how well SimpleTrack predicts $m$. Recall that the training is based on predicting the minimum $m$ needed for a reconstruction error of 50cm. For each test segments, we compute the actual $m$ needed to achieve 50cm reconstruction error; the actual $m$ needed is used as the reference. Figure \ref{fig:adaptiveVSnonAdaptive} shows the percentage error in predicting the actual $m$ needed to achieve 50cm reconstruction accuracy. It shows that the SVR-based algorithm gives accurate prediction.

\RR{
\subsection{System Performance}
\label{sec:sysperf}
We implemented the SimpleTrack encoding process on a real world sensor platform to understand its resource requirements. We use the Tmote Sky sensor node platform \cite{datasheet2005moteiv}, which features an 8MHz Texas Instruments MSP430 microcontroller with 10 kB Random Access Memory (RAM) and 48 kB flash programmable Read-Only Memory (ROM), as our hardware test environment. We use Contiki \cite{dunkels2004contiki} as our software test environment. Contiki is a C-based cooperative multi-threaded operating system for WSNs.}

\RR{
We use the animal dataset which has a higher number (128) data points in each trajectory segment. The maximum number of projections $m_{\max}$ needed for this dataset is 47. In principle, we need two different projection matrices, one for Northing and Easting. However, we find that using one projection matrix for both Northing and Easting can give us the same ADE so our implementation uses only one projection matrix. The dimension of the projection matrix $\tilde{U}$ (see Algorithm~\ref{alg:simpleTrackEncoding}) that a sensor node needs to store in this case is $47 \times 128$, which requires 24kB of memory storage. We also use a relatively large look-up table with 100 data pairs to stress the system. Table~\ref{tab:memory_power_acat} shows the computation time and energy consumption of the one look-up and one projection computation. Since each trajectory requires two projection matrix calculations (for northing and easting) and one look-up, the worst computation time (when 47 projections are needed) is $2 \times 47 \times 1.86 + 22.27 = 197.11ms$. Comparing to the duration of a trajectory, which is $128 \times 0.5 = 64s$, the computation time is less than 1\% of the time of a segment. Finally, the power consumption is based on the method described in~\cite{osterlindcontiki}. 
}


\begin{table}[t]
\centering
\caption{\RR{Memory and energy consumption of SimpleTrack.}}
\resizebox{!}{.6cm}{
\setlength{\tabcolsep}{1pt}
\begin{tabular}{|l|r|r|r|r|} \hline
Process&Current (mA) &Time (ms) &Voltage (V)&Power (mW) \\
\hline
Projection& 19.5 &$1.86$&3&0.10\\
\cline{1-5}
Lookup& 19.5 &$22.27$&3&1.30\\\hline
\end{tabular}
}
\label{tab:memory_power_acat}
\end{table}

\section{Related Work}
\label{sec:related}

The solution provided in this paper spans two key aspects: (1) Adaptive compression; (2) Deterministic construction of the projection matrix. We will therefore discuss the literature in these two areas. In addition, we discuss trajectory compression algorithms for WSNs.

\subsection{Adaptive Compression in WSNs}
Most adaptive compression algorithms proposed for WSN are motivated by energy savings. Most of the them considered slowly changing natural phenomena, which intrinsically require relatively low sampling. For example, \cite{1208942} proposed an adaptive compression algorithm, wherein compression is adapted at the sensing node by analyzing the correlation in a centralized data store. Since the approach requires central server to node communication, it is suitable for slowly changing phenomena e.g., soil moisture. However, we consider trajectory sampled at as high as 2 Hz sampling rate, therefore, such technique may result in enormous node to base communication causing quick depletion of the sensor node battery. 


Some other adaptive compression algorithms, although not requiring enormous inter-node communication, however require a large number of on-node processing.  For example, \cite{dong2006adaptive} proposed an adaptive wavelet compression algorithm for WSNs. In the proposed method each receiving sensor computes the space savings, and calculates the total energy dissipation to make a decision about whether to adjust the wavelet transform level. 
Clearly, this method will involve enormous computation given that for each trajectory segment it has to iterate multiple times to determine the optimal transform level for the best compression and energy trade-off.

A similar problem will be experienced with the algorithm proposed in~\cite{puthenpurayil2007energy}, which employs a feedback approach in which the space savings is compared to a pre-determined threshold. The compression model used in the previous frame can be retained and used for the next frame, if space savings is greater than the predefined threshold. Otherwise, the system will produce a new compression model. 

A slightly different adaptive compression principle is proposed in \cite{akyildiz2006state}. The authors design an on-line adaptive algorithm that dynamically makes compression decisions to accommodate the changing state of WSNs. 
By using the queueing model, the algorithm predicts the compression effect on the average packet delay and performs compression only when it can reduce the packet delay. 

\subsection{Deterministic Construction of Projection Matrix}
Random projection matrices are often used in compressive sensing because they are easy to generate and have provably good performance. However, random projection matrices may give rise to variability in performance \cite{Shen:2013hj}. This problem can be overcome by using deterministic construction of projection matrices. This gives rise to some well known methods to compute optimised projection matrices, see \cite{elad2007optimized,duarte2009learning,Carin:12}. These methods have already been discussed in the main text so we will not repeat the discussion here. A main contribution of this paper is a new method to compute a good projection matrix from dictionary. 


\subsection{Trajectory/GPS Compression Algorithms for Wireless Sensor Networks}
A very small number of papers can be found in the literature wherein trajectory compression algorithm for WSNs or other embedded platforms have been proposed. An example is~\cite{dttc}. This algorithm performs recursive 
segmentation of the trajectory, 
until a trajectory segment can be modeled with an interpolation function with a small error. 
Compression is achieved by only transmitting 
the relevant parameters of the interpolation function. 
However, computation requirement of this compression algorithm is high.

In~\cite{ghica2010trajectory} the authors propose a trajectory compression algorithm which uses various line simplification methods, for example, Dead-Reckoning and the Douglas-Peuker algorithm, and a variant of a CG-based optimal algorithm for polyline reduction. In particular, the authors also propose a hybrid approach, which combines some of the above methods. Note that out of the three methods, Douglas-Peuker is most popular. In our previous work~\cite{Rana:2011:AAC:1966251.1966255}, we have already shown that the non-optimized version of compressive sensing based trajectory compression method (the DCT-G method in Section \ref{sec:eval}) performed better than the improved Douglas-Peuker method proposed by Meratina et al~\cite{meratnia2004spatiotemporal}. In this paper, we show that our proposed method SimpleTrack outperforms DCT-G, SQUISH and a number of other methods.



\section{Conclusions}
\label{sec:conclusion}
Motivated by applications in mobile sensor networks, we investigate trajectory compression on resource constrained sensor nodes. We propose an adaptive trajectory compression method which is based on compressive sensing. Our proposed method has three key aspects. First, it uses dictionary learning. Second, it uses a new method to compute a good projection matrix from the dictionary. Third, it uses an adaptive algorithm to determine the number of projections required. We evaluate the performance of our proposed method using 6 datasets. We show that our algorithm outperforms many existing algorithms. In fact, our algorithm can achieve sub-metre accuracy. The algorithm also has low computation requirements and can be used on resource impoverished sensor platforms. 

\ifCLASSOPTIONcaptionsoff
  \newpage
\fi




\begin{spacing}{0.85}
\bibliographystyle{IEEEtran}
\bibliography{reference_3,sigproc,csbib}
\end{spacing}

\end{document}